\documentclass[twocolumn]{aastex7}
\usepackage{enumitem} % 在导言区添加
\usepackage{graphicx}
\usepackage{CJK}
\usepackage{lineno}
\usepackage{amsmath}
\usepackage{float}
\usepackage{tabularx}
\submitjournal{xxx}
\begin{document}
\begin{CJK*}{UTF8}{gbsn}
\title{Joint Analysis of Optical, Near-Infrared And Mid-Infrared Variability of 4 Quasars at Redshift $<$ 1
%in the CatNorth and VVV/VVVX Catalogues
}
\shortauthors{Long et al.}
\correspondingauthor{Zhen-Ya Zheng}
\email{zhengzy@shao.ac.cn}

\author[orcid=0000-0003-3787-0790]{Lin Long({\CJKfamily{gbsn}龙琳})}
\affiliation{Shanghai Astronomical Observatory, Chinese Academy of Sciences, 80 Nandan Road, Shanghai, 200030, People's Republic of China}
\affiliation{School of Astronomy and Space Sciences, University of Chinese Academy of Sciences, No. 19A Yuquan Road, Beijing 100049, People's Republic of China}
\email{longlin@shao.ac.cn}  

\author[orcid=0000-0002-9634-2923]{Zhen-ya Zheng({\CJKfamily{gbsn}郑振亚})} 
\affiliation{Shanghai Astronomical Observatory, Chinese Academy of Sciences, 80 Nandan Road, Shanghai, 200030, People's Republic of China}
\email{zhengzy@shao.ac.cn}

\author[orcid=0000-0002-7152-3621]{Ning Jiang({\CJKfamily{gbsn}蒋凝})} 
\affiliation{Department of Astronomy, University of Science and Technology of China, Hefei 230026, People's Republic of China}
\email{jnac@ustc.edu.cn}

\author{Chun Xu({\CJKfamily{gbsn}许春})} 
\affiliation{Shanghai Astronomical Observatory, Chinese Academy of Sciences, 80 Nandan Road, Shanghai, 200030, People's Republic of China}
\email{chun.xuu@shao.ac.cn}

\author[orcid=0000-0002-3134-9526]{Jiaqi Lin({\CJKfamily{gbsn}林家琪})}
\affiliation{School of Physics and Astronomy, Sun Yat-sen University, Zhuhai 519082, People's Republic of China}
\affiliation{Shanghai Astronomical Observatory, Chinese Academy of Sciences, 80 Nandan Road, Shanghai, 200030, People's Republic of China}
\affiliation{CSST Science Center for the Guangdong-Hong Kong-Macau Greater Bay Area, Sun Yat-sen University, Zhuhai 519082, People's Republic of China}
\email{linjiaqi@shao.ac.cn}

\author[orcid=0000-0001-6763-5869]{Fang-Ting Yuan({\CJKfamily{gbsn}袁方婷})}
\affiliation{Shanghai Astronomical Observatory, Chinese Academy of Sciences, 80 Nandan Road, Shanghai, 200030, People's Republic of China}
\email{xxx}

\author[orcid=0000-0002-0003-8557]{Chunyan Jiang({\CJKfamily{gbsn}姜春艳})}
\affiliation{Shanghai Astronomical Observatory, Chinese Academy of Sciences, 80 Nandan Road, Shanghai, 200030, People's Republic of China}
\email{xxx}

\author[0000-0003-3987-0858]{Ruqiu Lin
({\CJKfamily{gbsn}林如秋})}
\affiliation{Shanghai Astronomical Observatory, Chinese Academy of Sciences, 80 Nandan Road, Shanghai, 200030, People's Republic of China}
\affiliation{School of Astronomy and Space Sciences, University of Chinese Academy of Sciences, No. 19A Yuquan Road, Beijing 100049, People's Republic of China}
\email{linruqiu20@mails.ucas.ac.cn}

\author[orcid=0000-0002-1530-2680]{Hai-Cheng Feng({\CJKfamily{gbsn}封海成})}
\affiliation{Yunnan Observatories, Chinese Academy of Sciences, Kunming 650216, Yunnan, People's Republic of China}
\affiliation{Key Laboratory for the Structure and Evolution of Celestial Objects, Chinese Academy of Sciences, Kunming 650216, Yunnan, People's Republic of China}
\affiliation{Center for Astronomical Mega-Science, Chinese Academy of Sciences, 20A Datun Road, Chaoyang District, Beijing 100012, People's Republic of China}
\affiliation{Key Laboratory of Radio Astronomy and Technology, Chinese Academy of Sciences, 20A Datun Road, Chaoyang District, Beijing 100101, People's Republic of China}
\email{hcfeng@ynao.ac.cn}

\author[0000-0001-8416-7059]{Hengxiao Guo({\CJKfamily{gbsn}郭恒潇})}
\affiliation{Shanghai Astronomical Observatory, Chinese Academy of Sciences, 80 Nandan Road, Shanghai, 200030, People's Republic of China}
\email{hengxiaoguo@gmail.com}

\author[orcid=0000-0002-0539-8244]{Xiang Ji({\CJKfamily{gbsn}吉祥})} 
\affiliation{Shanghai Astronomical Observatory, Chinese Academy of Sciences, 80 Nandan Road, Shanghai, 200030, People's Republic of China}
\email{jixiang@shao.ac.cn}

% \author{et al.} 
% \affiliation{xx}
% \email{xx}

% \collaboration{all}{The Terra Mater collaboration}

\begin{abstract}
Amid rapid advances in time-domain astronomy, multi-wavelength (e.g., optical and infrared) time-domain studies of quasars remain scarce. Here we present a systematic analysis of four quasars initially selected by their Ks-band variability amplitudes in the VISTA Variables in the V\'{\i}a L\'actea Survey (VVV/VVVX). For these objects, we obtain complementary optical light curves from Pan-STARRS1 (PS1) and the Zwicky Transient Facility (ZTF), and W1-band light curves from the Wide-field Infrared Survey Explorer (WISE). We perform correlation analysis to study the time lags between different bands, which may be directly related to the size of the dust torus. After correcting for infrared flux contamination from the accretion disk and accounting for the redshift effect, we measure the Ks-optical and W1-optical lags for the targets VVV J1834-2925 and VVV J1845-2426. Using typical sublimation temperatures and reverberation time lags, we obtain a graphite-to-silicate grain size ratio of 
$\frac{a_C}{a_S}\sim$ 0.4. Through SED fitting, we determine the luminosities of these quasars and find that their dust torus sizes follow the established $R_{dust}-L_{AGN}$ relation reported in previous studies.
\end{abstract}

% \keywords{\uat{Galaxies}{573} --- \uat{Cosmology}{343} --- \uat{High Energy astrophysics}{739} --- \uat{Interstellar medium}{847} --- \uat{Stellar astronomy}{1583} --- \uat{Solar physics}{1476}}


\section{Introduction} 
Active Galactic Nuclei (AGNs) are among the brightest objects in the universe, powered by the accretion of matter onto a central supermassive black hole (SMBH). As matter falls into the black hole's gravitational well, it releases vast amounts of energy, producing intense radiation across the electromagnetic spectrum, ranging from radio band to gamma rays. The canonical model of an AGN \citep{1993ARA&A..31..473A, 1995PASP..107..803U} describes a central SMBH surrounded by an accretion disk, with emission lines originating in the broad-line region (BLR) close to the black hole and the narrow-line region (NLR) at a 
further distance. Obscuration along equatorial lines of sight is attributed to a circumnuclear, toroidal distribution of dust (dusty torus), which determines the Type 1/Type 2 dichotomy of AGN.

The dust torus absorbs the UV/optical radiation emitted from the accretion disk and re-emits it as infrared radiation \citep{1978ApJ...226..550R}, with the light-crossing time of the torus introducing a measurable lag between optical emission and IR echo. From the perspective of spatial distribution of radiation, the dust torus extends outwards from the dust sublimation boundary of the BLR \citep{1987ApJ...320..537B}. The dust compositions in the torus such as graphite and silicate grains exhibit distinct thermal attributes, with maximum sublimation temperatures of 1800 K and 1400 K, respectively \citep{2015ARA&A..53..365N}. These critical temperatures define the survival zone of dust, establishing the sublimation radius as the inner boundary for stable dust existence and participation in radiative processes.
%As the temperature of different torus regions varies with distance from the central engine, their radiation characteristics show clear spatial differentiation. The near-infrared (NIR) emission primarily originates from the hot inner edge of the torus (near the sublimation radius), while mid- (MIR) to far-infrared (FIR) radiation arises from cooler outer regions \citep{2008ApJ...685..160N, 2015ARA&A..53..365N}. This temperature gradient results in a continuous infrared spectral energy distribution (SED) in AGNs, spanning from NIR to FIR wavelengths, reflecting the multi-layered thermal structure of the torus.

For nearby Seyfert galaxies, infrared (IR) interferometry can directly probe the geometry of their nuclear IR structures on parsec scales. These studies based on IR interferometry reveal significant mid-infrared (MIR) emission from the polar regions of AGNs (e.g., \citealt{2000AJ....120.2904B, 2005ApJ...618L..17P, 2010A&A...515A..23H, 2010MNRAS.402..879R, 2016A&A...591A..47L, 2018ApJ...862...17L, 2022A&A...663A..35I, 2023A&A...678A.136I, 2022Natur.602..403G}). \cite{2012ApJ...755..149H, 2013ApJ...771...87H} proposed a radiatively-driven dust wind model that provides a qualitative explanation for this phenomenon. According to this model, the polar dust originates from an optically thin dusty outflow, launched from the hottest, innermost regions of the optically thick dusty torus by radiation pressure. This wind extends parsec- to tens of parsec-scale distances along the polar direction. In this scenario, the near-infrared (NIR) emission is dominated by the torus, while the MIR emission primarily arises from the dusty outflow.

\citet{2018ApJ...866...92L} characterized the IR spectral energy distribution (SED) features of polar dust emission in AGNs based on successful SED fitting of 64 low-redshift Seyfert-1 nuclei. Their results indicate that the infrared SEDs of most AGNs originate from a similar circumnuclear torus structure, with variations primarily caused by differences in the optical depth of an extended obscuring dust component. Furthermore, the similar far-infrared (FIR) SED shapes observed across diverse AGN types suggest that this emission likely originates from an optically thin and spatially extended dust structure \citep{2017ApJ...841...76L}, most likely in the form of extensive polar dust.

Quasars, a particularly bright type of AGN, appear as extremely luminous point-like sources with high luminosity, high redshift, and significant brightness variability \citep{1997iagn.book.....P}. The intrinsic variability of AGNs \citep{1997ARA&A..35..445U} offers an unique opportunity to investigate their internal structure and physical mechanisms through techniques such as reverberation mapping (RM). RM observations have been conducted in combinations of UV/optical-NIR (e.g., \citealt{1974MNRAS.169..357P, 1992MNRAS.256P..23G, 2004MNRAS.350.1049G, 2004ApJ...600L..35M, 2019ApJ...886..150M, 2009ApJ...700L.109K, 2014ApJ...788..159K}), UV/optical-MIR (e.g., \citealt{2015ApJ...801..127V, 2019ApJ...886...33L, 2020ApJ...900...58Y, 2025arXiv250421711S}) and UV/optical-NIR-MIR (e.g., \citealt{2013sptz.prop10162W, 2018rnls.confE..57S, 2021ApJ...912..126L}). 

Nevertheless, RM studies that simultaneously cover optical, NIR, and MIR bands remain rare, which are limited in sample size compared to analyses focusing only on optical-NIR or optical-MIR wavelengths. By combining optical and NIR-MIR RM data, \citet{2018rnls.confE..57S} revealed a bowl-shaped geometry of the dust torus in the nearby Seyfert 1 galaxy WPVS 48 and clearly distinguished the spatial distribution of the BLR from that of the dust torus. \citet{2021ApJ...912..126L} performed optical-NIR-MIR RM of NGC4151 based on 24-year light curves from the Crimean Observatory of the Sternberg Astronomical Institute (SAI) monitoring, which unveiled a two-component dust torus structure and detected a 4$\%$ yr$^{-1}$ increasing trend in hot dust emission, suggesting possible physical growth of the torus. 

Current multi-band (optical-NIR-MIR) 
RM studies are constrained by the lack of large-scale NIR time-domain surveys. As a result, NIR data are primarily obtained through targeted observations with specific telescopes, thereby limiting applications to a small number of known, highly variable AGNs. While this approach is well-suited for in-depth case studies of exceptional targets, it hinders the construction of statistical samples. The current limitations highlight the need to combine NIR, optical, and MIR survey data for systematic identification of strongly variable AGNs across multiple bands, which would significantly enhance the sample size for RM studies. In addition to normal AGN variability, large amplitude of flares in AGNs, identified as TDEs or changing-look AGNs also provide us an ideal opportunity. Particularly, the optical-IR time lag of these flares can be easily measured.

Based on the constraints of dust torus size derived from RM time lag measurements, researchers further investigated the radius-luminosity ($R_{dust}-L$) relation. Through a series of RM observations spanning optical to NIR and MIR, \cite{2006ApJ...639...46S, 2014ApJ...788..159K, 2019ApJ...886..150M, 2019ApJ...886...33L, 2020ApJ...900...58Y, 2023MNRAS.522.3439C, 2024ApJ...968...59M} statistically established the $R_{dust}-L$ relation. They found that the dust torus size exhibits a positive correlation with AGN luminosity, following a power-law relationship with an index close to 0.5, consistent with predictions from the dust sublimation model \citep{1987ApJ...320..537B}. However, observations \citep{2001ASPC..224..149O, 2007A&A...476..713K, 2008ApJ...685..147N} also revealed that the actual dust torus sizes are smaller than the prediction of the dust sublimation model. To explain this discrepancy, \citet{2011ApJ...737..105K, 2014ApJ...797...65W, 2023MNRAS.522.3439C} proposed that self-shadowing effects of the slim disk might play a role, and \citet{2019ApJ...886..150M} suggested that the anisotropic radiation from the accretion disk leads to a smaller dust sublimation radius along the equatorial plane than in the polar direction. Additionally, \citet{2019ApJ...886..150M} noted that the observed prevalence of larger dust torus radii in low-luminosity AGNs is due to thermal lag effects in the thermal lag of the dust's response to luminosity variations. Thus, the investigation of AGN dust tori constitutes a complex challenge. A comprehensive analysis will require the unique capabilities of now-operational facilities: high-cadence optical monitoring from the Rubin Observatory \citep{LSST-2019ApJ} to track the driving continuum, and complementary wide-field infrared spectroscopy from the Spectro-Photometer for the History of the Universe, Epoch of Reionization, and ices Explorer \citep[SPHEREx;][]{2020SPIE11443E..0IC} to probe the dust response.

% The cross-correlation function (CCF), as a conventional method for measuring time lags, essentially determines the time delay corresponding to the maximum correlation by computing the correlation between two light curves at different temporal offsets. Its main advantage is that the method is simple and fast. The Damped Random Walk (DRW) model is an effective statistical model used to describe AGNs's light curves on timescales ranging from months to years. It characterizes AGNs variability as a stochastic process with an exponential decay in correlation over time. Currently, several popular implementations, such as \emph{Javelin} \citep{2011ApJ...735...80Z, 2013ApJ...765..106Z, 2016ApJ...819..122Z} and \emph{taufit} \citep{2021Sci...373..789B, 2024ApJ...975..160R}, are available as software packages that utilize DRW model to model light curves. For sparse light curves, cross-correlation method struggles to measure lags directly. \emph{Javelin} improves this by statistically combining multiple light curves, modeling their correlations, and removing mean trends \citep{2011ApJ...735...80Z, 2013ApJ...765..106Z, 2016ApJ...819..122Z}. \zzy{JN: I dno't think this paragraph is necessary to appear in the introduction}

In this work, we perform torus lag analysis on four quasars. 
In Section \ref{sec:data}, we describe the data and the construction of multiband light curves.
We introduce two RM techniques and report the measured NIR and MIR lag times in 
Section \ref{sec:analysis}. We present the analysis of the structure function in Section \ref{sec:sf}. In Section \ref{sec:dust}, we analyze the dust properties and examine the relationship between dust torus size and AGN luminosity. Finally, we provide our conclusion in Section \ref{sec:highlight}. Throughout this study, we adopt $H_{0}$=70~$\rm{km\ s^{-1}\ Mpc^{-1}}$, $\rm{\Omega}_{m}=0.3$ and $\rm{\Omega_{\Lambda}=0.7}$.

%% The "ht!" tells LaTeX to put the figure "here" first, at the "top" next
%% and to override the normal way of calculating a float position.
%% The asterisk after "figure" tells the compiler to span multiple columns
%% if a two column style is selected.

\begin{table*}
\begin{center}
\caption{The four quasars' basic information and light curves used in this work.
\label{table:lc}
}
\renewcommand{\arraystretch}{1.5}
\begin{tabular*}{\linewidth}{@{\extracolsep{\fill}}cccccccccccc@{}}
 \cline{1-12}
 Name & RA & DEC & z$_{ph}$ & z$_{spec}$ & Ks & W1 & W2 & SNR(Ks) & Survey & Filter & Time Span \\
 (1) & (2) & (3) & (4) & (5) & (6) & (7) & (8) & (9) & (10) & (11) & (12)
 \\
 \cline{1-12} 
 VVV J1831-2714 & 277.75018 & -27.23505 & 0.847 & - & 15.059 & 12.400 & 12.178  & 1383.4 & PS1 &  g, r, i, z, y & 2010 $\sim$ 2014
 \\
 &&&&&&&&& VVVX & Ks & 2010 $\sim$ 2015
\\ 
&&&&&&&&& WISE & W1, W2 & 2014 $\sim$ 2024
\\
 \cline{1-12}
 VVV J1834-2925 & 278.52821 & -29.42018 
    & 0.936 & - & 14.304 & 12.239 & 11.384 & 481.2
    & PS1 
    & g, r, i, z, y
    & 2010 $\sim$ 2014
\\
&&&&&&&&& VVVX & Ks & 2010 $\sim$ 2019
\\
&&&&&&&&& WISE & W1, W2 & 2014 $\sim$ 2024
 \\ 
 \cline{1-12}
 VVV J1833-2731 & 278.40347 
    & -27.51866 
    & 0.275 & - & 15.229 & 13.653 & 12.664 & 238.0
    & PS1 
    & g, r, i, z, y
    & 2009 $\sim$ 2014
\\
&&&&&&&&& ZTF & g, r & 2018 $\sim$ 2024
\\
&&&&&&&&& VVVX & Ks & 2010 $\sim$ 2017
\\
&&&&&&&&& WISE & W1, W2 & 2014 $\sim$ 2024
 \\ 
 \cline{1-12}
 VVV J1845-2426 & 281.30173 
    & -24.44572 
    & 0.308 & 0.2547 & 13.686 & 12.123 & 11.033 & 425.6
    & PS1 
    & g, r, i, z, y
    & 2009 $\sim$ 2014
 \\
&&&&&&&&& ZTF & g, r & 2018 $\sim$ 2024
\\
&&&&&&&&& VVVX & Ks & 2010 $\sim$ 2017
\\
&&&&&&&&& WISE & W1, W2 & 2014 $\sim$ 2024
 \\ 
 \cline{1-12}
\end{tabular*}
\end{center}
\raggedright Notes: Column (1): identifier of the quasar from the VVV survey. Columns (2)-(3): equatorial coordinates of the quasar (units: degree). Column (4): ensemble photometric redshift from \cite{2024ApJS..271...54F}. Column (4): The spectral redshift calculated by LJT spectrum. Columns (6)-(8): average magnitudes of the Ks, W1 and W2 band light curves (units: mag). Column (9): The SNR of the Ks-band light curve, which is calculated by [max(Ks)$-$min(Ks)]$^2$/2$\bar{\sigma}^2$ (refer to Sec \ref{sec:data sample}). Columns (10)-(11): survey name and corresponding filter band. Column (12): observation period of the light curve.
\end{table*}

\section{Data} \label{sec:data}
\subsection{Sample}\label{sec:data sample}
To investigate internal structures of quasars, we select targets from the quasar candidate catalog of \cite{2024ApJS..271...54F}, combining them with available optical and infrared light curves. \cite{2024ApJS..271...54F} constructed an improved quasar candidate catalog, CatNorth, based on data from Gaia, Pan-STARRS1, and CatWISE2020, as an enhanced version of the Gaia Data Release 3 (Gaia DR3) quasar candidate catalog. By applying the XGBoost algorithm on the parent Gaia DR3 catalog, they have identified 1,545,514 high-confidence quasar candidates with photometric redshifts. 
The CatNorth catalog achieves $\sim$90$\%$ high purity while maintaining excellent completeness. 

We cross-match CatNorth with the VISTA Variables in the V\'{\i}a L\'actea Survey (VVV) and its extension (VVVX) and identify 55 common sources. %, among which only 4 sources exhibited significant variability ($\Delta mag > 0.5 mag$) in the Ks-band time series data. 
After extracting the Ks-band light curves of these sources from VVV/VVVX data, we calculate their signal-to-noise ratios (SNR), SNR=[max(Ks)$-$min(Ks)]$^2$/2$\bar{\sigma}^2$, where $\bar{\sigma}$ is the mean error of the Ks-band light curve. The SNRs of light curves are required to be not lower than 200 to ensure data quality sufficient for RM analysis. Applying the above cut in SNR produced a final quasar sample of four objects. All four quasars in our sample show $\Delta Ks > 0.5$ mag. The redshifts of these four quasars are all less than 1 \citep{2024ApJS..271...54F}, and their basic properties are listed in Table \ref{table:lc}.
%\zzy{How significant variability on selecting these four quasars? Basic information of the 4 quasars, such as name, redshift, brightness, etc.}
% \zzy{JN: This part lacks detail. Why did you choose 0.5 mag variability? How did you calculate it? What time range did you use? }

\begin{figure*}[!ht]
\centering
\includegraphics[width=0.95\linewidth]{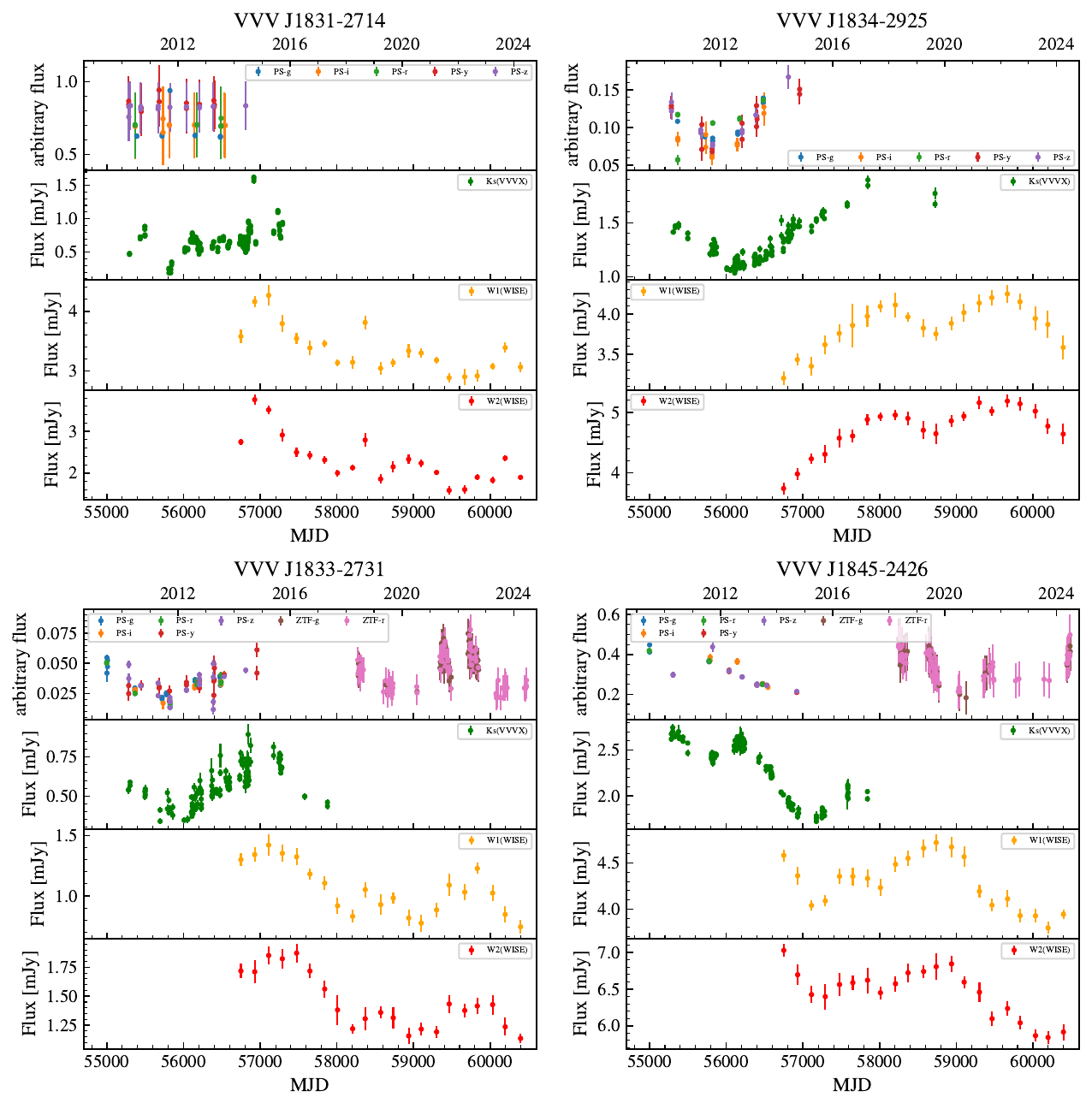}
\caption{The optical, NIR and MIR light curves of four quasars between 2010 and 2024. These were collected from PS1 (g, r, i, z, y), ZTF (g, r), VVV/VVVX (Ks) and WISE (W1, W2). The optical light curves were calibrated by PyCALI.\label{lc}}
\end{figure*}

\subsection{Optical and Infrared Time-domain Surveys}
% \zzy{JN: I don't think WISE is formally a transient survey, maybe "Time-domain" should be a more accurate term. A transient survey should discover transients and issue alerts in real-time, which the WISE survey does not do. }
In this study, we focus on the analysis of the four variable quasars with their optical and infrared lightcurves from Pan-STARRS1 (PS1), the Zwicky Transient Facility (ZTF), VVV/VVVX and the Wide-field Infrared Survey Explorer (WISE). These surveys are introduced briefly as below.

PS1\footnote{\url{https://catalogs.mast.stsci.edu/panstarrs/}} conducted a series of multi-object astronomical surveys between May 2010 and March 2014, using the 1.8-meter telescope and a five-band filter system (g$\sim 4810$ \AA, r$\sim 6170$ \AA, i$\sim 7520$ \AA, z$\sim 8660$ \AA, y$\sim 9620$ \AA) \citep{2012ApJ...750...99T, 2020ApJS..251....7F}.

ZTF\footnote{\url{https://www.ztf.caltech.edu/}} employs the Palomar 48-inch Schmidt
Telescope (P48) to scan the entire northern visible sky  with a 47 deg$^2$ field of view (FOV) to median depths of g (4784 \AA) $\sim$ 20.8, r (6417 \AA) $\sim$ 20.6 and i (7867 \AA) $\sim$ 19.9 mag (AB), respectively \citep{2019PASP..131g8001G, 2019PASP..131a8003M}. We obtain time series data in the g and r bands from the ZTF Data Release 23 for two quasars. 

VVV is a near-infrared, multi-epoch photometric survey conducted using the VISTA telescope. From 2009 to 2015, it observed 560 deg$^2$ of the Milky Way's bulge and southern disk, including approximately 100 epochs in the Ks band and additional observations in the Z, Y, J, and H bands \citep{2010NewA...15..433M}). Its extension, the VVVX survey, expanded the coverage to 1,700 deg$^2$ between 2016 and 2023, adding more epochs in the J, H, and Ks bands and extending the time baseline by more than a factor of two \citep{2024A&A...689A.148S}. We use light curve in the Ks band from VVV DR5\footnote{\url{http://vsa.roe.ac.uk/}} and VVVX\footnote{\url{https://archive.eso.org/}}.

WISE conducted an all-sky survey in four infrared bands centered at 3.4, 4.6, 12, 22 $\mu m$ (W1, W2, W3 and W4) from December 2009 to September 2010 \citep{2010AJ....140.1868W}. Subsequently, the Near-Earth Object Wide-field Infrared Survey Explorer Reactivation (NEOWISE) mission observed in the W1 and W2 bands for four months to search for asteroids within the solar system until its mission concluded in February 2011 \citep{2011ApJ...731...53M}. NEOWISE observations resumed in December 2013 to further search for asteroids and comets that could pose an impact threat to Earth, a mission known as NEOWISE-Reactivation (NEOWISE-R), which ended in August 2024 \citep{2014ApJ...792...30M}. 

\subsection{Construction of Multiband Light Curves}
We have optical, near-infrared and mid-infrared band light curves from PS1 (g, r, i, z, y), ZTF (g, r), VVVX (Ks) and WISE (W1, W2). For the PS1 data, we utilize the catalog's "psfFlux" measurements. The ZTF magnitudes were converted to flux using zeropoints from the Spanish Virtual Observatory (SVO) Filter Profile Service\footnote{\url{http://svo2.cab.inta-csic.es/theory/fps/index.php?mode=browse&gname=Palomar&gname2=ZTF&asttype=}} \citep{2024ApJ...968...59M}. The optical observations are derived from two different surveys, each using different telescopes and filters. To build a unified light curve, we inter-calibrate the data to ensure consistent flux scaling. For this purpose, we use the PyCALI\footnote{\url{https://github.com/LiyrAstroph/PyCALI}} algorithm, which models the light curves with a Damped Random Walk (DRW) process and employs a Bayesian framework to obtain the best scaling parameter during inter-calibration \citep{2024zndo..10700132L}. The calibrated optical light curves are shown in Figure \ref{lc}.

We convert VVV/VVVX's "APERMAG3" (Vega magnitude system) into flux (in units of Jy) using $m_{vega}+\Delta m = -2.5log_{10}f_{\nu}-48.6$. The $\Delta m$ is the conversion term from the Ks-band Vega magnitude system to the AB magnitude system, which is 1.839 \footnote{\url{http://casu.ast.cam.ac.uk/surveys-projects/vista/technical/filter-set}}. 

We reject the low-quality observations in the NEOWISE data using the following criteria: "cc$\_$flags$\neq$0000", "nb$>$1", "na$>$0", "pf$\_$qual$\neq$AA" and "qual$\_$frame=0". Observations were discarded if any of the above conditions were satisfied. The Vega magnitudes of the remaining data points were adjusted to the AB system by applying offsets ($\Delta m$) of 2.699 (W1) and 3.339 (W2), followed by flux transformation \citep{2011ApJ...735..112J}. We average the flux measurements in the W1 and W2 bands over 30-day epochs. The flux error for a given epoch was calculated as the quadrature sum of: (1) the standard deviation of fluxes within that epoch, (2) the photometric measurement errors ($\sigma_{i,pho}$), and (3) the system stability uncertainties ($\sigma_{s.s.}$). So the flux and flux error for a given epoch can be described as 
% \begin{center}
\begin{equation}
f_{epoch} = \frac{1}{N}\sum^{N}_{i=1}f_i
\end{equation}
% \end{center}

\begin{center}
\begin{equation}
\sigma^2_{epoch}=\frac{1}{N-1}\sum^N_{i=1}(f_i-\overline{f_{epoch}})^2+\frac{1}{N^2}\sum^N_{i=1}\sigma^2_{i,pho}+\frac{1}{N}\sigma^2_{s.s.}.
\end{equation}
\end{center}
The light curves of these four quasars are shown in Figure \ref{lc} and Table \ref{table:lc}.

\subsection{Spectroscopic Redshift and Black Hole Mass}
We perform a spectroscopic observation of VVV J1845-2426 on August 27, 2025 (MJD 60914), using the 2.4-meter optical telescope (LJT) at the Lijiang Observatory \citep{2019RAA....19..149W}. The observation employs Grism 3, which provides a wavelength coverage of 3400-9100 $\AA$, a dispersion of 2.9 $\AA$ pixel$^{-1}$, and a slit width of 2.5$^{"}$. We conduct an exposure of 1800s on the target source. After obtaining the raw data, we subsequently perform spectral extraction using the \emph{PyFOSC} pipeline \citep{yuming_fu_2024_10967240}, which is a software written in Python specifically designed for reducing long slit spectral data. The redshift of VVV J1845-2426 is 0.2547, listed in Table \ref{table:lc}. It is noteworthy that, given the acquisition of a spectrum for VVV J1845-2426, all subsequent redshift-dependent calculations for this source employ its spectroscopic redshift, whereas photometric redshifts are utilized for the other sources lacking spectroscopic observations.

We use the Python package \emph{QSOFITMORE} \citep{2021zndo...5810042F}, which is built upon \emph{PyQSOFit} \citep{2018ascl.soft09008G, 2019ApJS..241...34S}, to fit the spectrum of our quasar. This code supports three Galactic extinction maps: "sfd" \citep{1998ApJ...500..525S}, "planck14" \citep{2014A&A...571A..11P}, and "planck16" \citep{2016A&A...596A.109P}. The fitting procedure of \emph{QSOFITMORE} includes: Galactic extinction correction, redshift correction, host galaxy component subtraction, quasar continuum fitting, and emission line fitting. The quasar continuum can be decomposed into a power law spectrum, an \ion{Fe}{2} template, and a polynomial component.

\begin{figure*}[!htbp]
\centering
\includegraphics[width=0.9\linewidth]{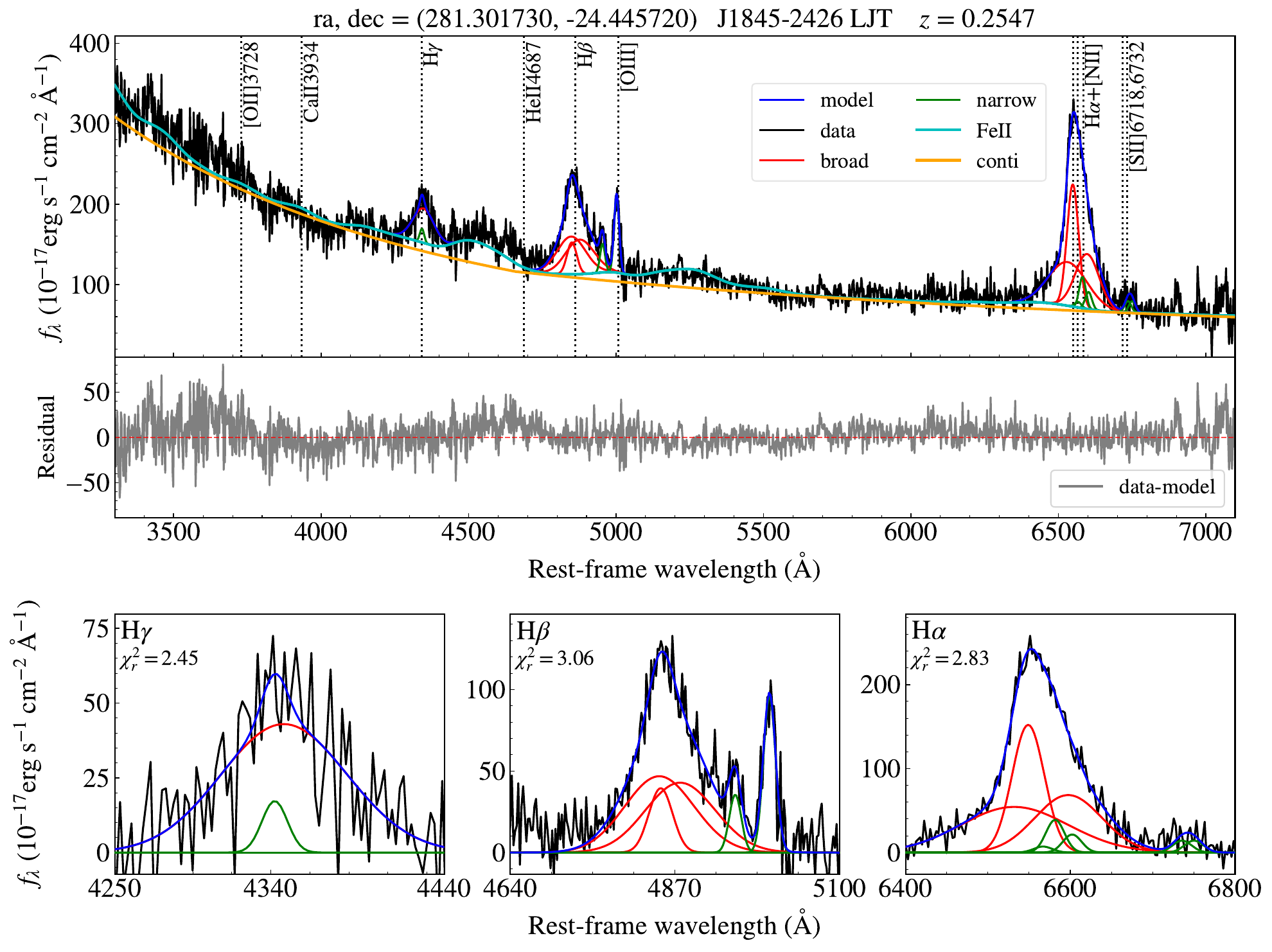}
\caption{Spectrum fitting results for VVV J1845-2426 (LJT). In the top panel, observed spectrum (black line) is decomposed into power law continuum (orange line), \ion{Fe}{2} (cyan line), and emission lines (blue line). The emission lines include the broad (red line) and narrow (green line) components. The middle panel displays the residuals between the data and the model. The bottom left and right panels show the fitting results 
for the $H_{\gamma}$, $H_{\beta}$ and $H_{\alpha}$ emission lines, respectively.
\label{line fitting}}
\end{figure*}

We perform a spectral fitting on VVV J1845-2426, shown in Figure \ref{line fitting}. Due to the strong AGN fraction on the observed spectrum, the host galaxy component is not subtracted in this case. As shown in the figure, we use three broad components to fit the broad $H_{\beta}$ line and a combination of three broad and three narrow components for the $H_{\alpha}$ line. Then, based on the fitting results of the broad $H_{\beta}$ and $H_{\alpha}$ emission lines, we estimated the mass of the central black hole \citep{2005ApJ...630..122G, 2006ApJ...641..689V}:
\begin{equation}\label{BH}
    \begin{split}
          \rm log(M_{BH}/{ M_\odot}) & = \rm 2 \  log(\frac{FWHM(H_ {\beta})}{km \ s^{-1}}) \\
           & \rm + 0.5 \ log(\frac{ L_{5100}}{10^{44} \ erg \ s^{-1}}) + 0.91 \\
    \end{split}
\end{equation}

\begin{equation}\label{BH2}
   \begin{split}
   \rm log(M_{BH}/{ M_\odot}) & = \rm 2.06 \  log(\frac{FWHM(H_ {\alpha})}{10^3 \ km \ s^{-1}}) \\
           & \rm + 0.55 \ log(\frac{ L_{H_{\alpha}}}{10^{42} \ erg \ s^{-1}}) + 6.3 \\
   \end{split}
\end{equation}

where $\rm M_{\odot}$ is the solar mass, FWHM($H_{\beta}$) and FWHM($H_{\alpha}$) are the full widths at half-maximum (FWHMs) of best-fitting broad $H_{\beta}$ and $H_{\alpha}$ emission line profiles (a composite of the three broad components). The L$_{5100}$ is the continuum luminosity at the rest frame 5100 $\AA$ and the L$_{H_{\alpha}}$ is calculated from the $H_{\alpha}$ broad emission line profile. The $\rm log(M_{BH}/M_{\odot})$ for VVV J1845-2426 are $8.9^{+0.1}_{-0.1}$ ($H_{\beta}$) and $8.3^{+0.2}_{-0.2}$ ($H_{\alpha}$). The errors of black hole mass are derived from the errors of $H_{\beta}$ and $H_{\alpha}$ line widths, respectively. The fiducial black hole mass is taken to be the average of the two measurements, yielding $\rm log(M_{BH}/M_{\odot})=8.6^{+0.1}_{-0.1}$. We note that this method of estimating black hole mass has a systematic uncertainty of 0.5 dex \citep{2013BASI...41...61S}. Some spectral fitting results and the mass of black hole are listed in Table \ref{table:spectral fitting}.

\begin{table}[htbp]
\centering % 用 \centering 代替 center 环境，它是更推荐的居中方式
\caption{Spectral Fitting Results for J1845-2426.}
\label{table:spectral fitting}
\renewcommand{\arraystretch}{1.5}
% 使用 tabularx 环境，总宽度设为 \linewidth (或 \textwidth)
% 三个 X 列会严格均分整个宽度
\begin{tabularx}{\linewidth}{@{} *{3}{>{\centering\arraybackslash}X} @{}}
 \hline
 \hline
  Parameter & H$_{\beta}$ & H$_{\alpha}$ \\
  \hline
  FWHM [km/s] & 5400$^{+530}_{-530}$ & 3210$^{+540}_{-540}$\\
  log(M$_{\mathrm{BH}}$/M$_{\odot}$) & 8.9$^{+0.1}_{-0.1}$ & 8.3$^{+0.2}_{-0.2}$ \\ % 使用 \mathrm 使 BH 和odot 正体显示
 \hline
\end{tabularx}
\end{table}

\begin{figure*}[!htbp]
\centering
\includegraphics[width=0.8\linewidth]{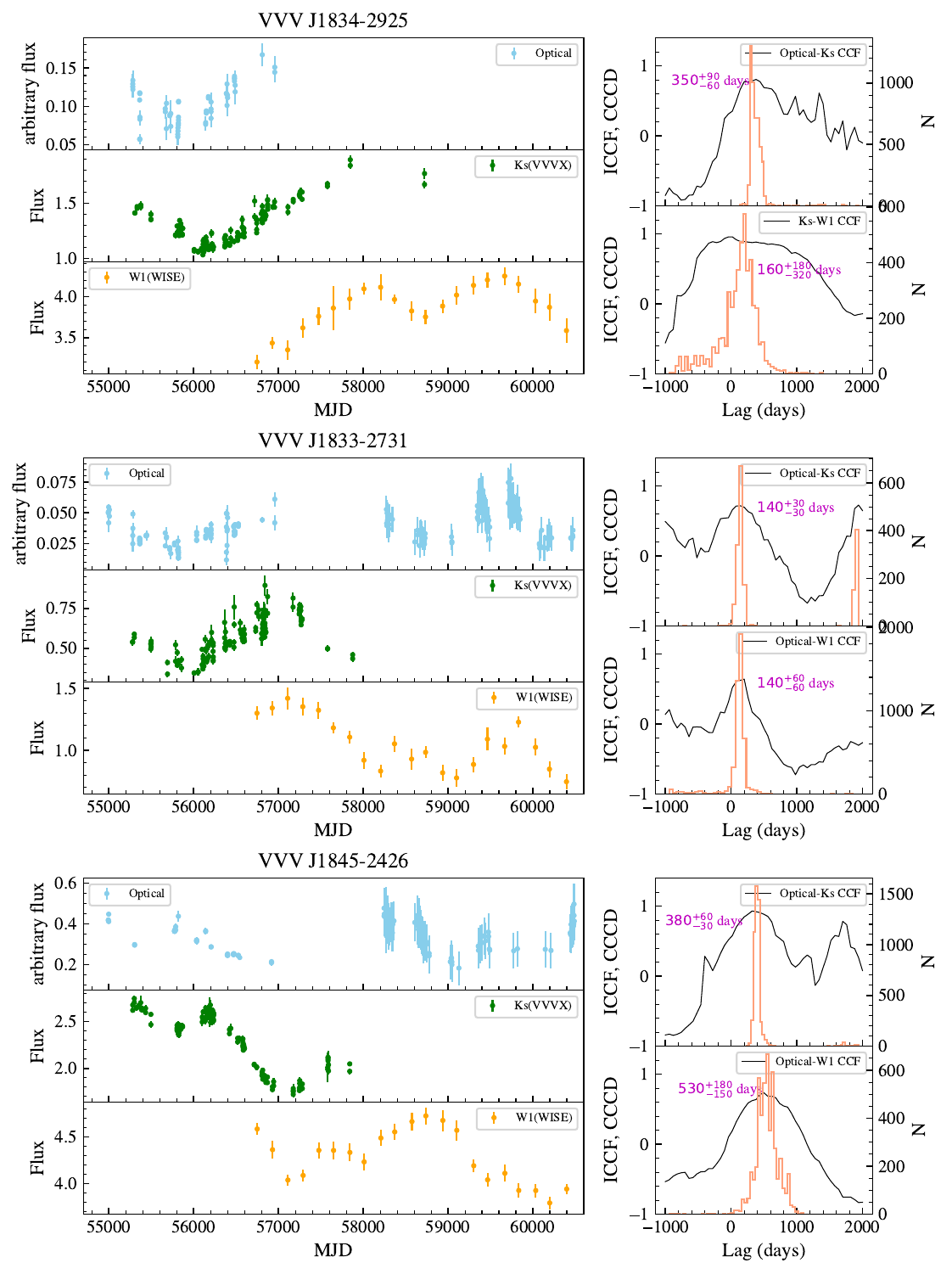}
\caption{Light curves (left) and ICCF results (right) for targets VVV J1834-2925 (top), VVV J1833-2731 (middle) and VVV J1845-2426 (bottom), respectively. The CCF analysis results in the upper right corner correspond to optical-NIR and NIR-MIR correlations, respectively. All other CCFs represent optical-NIR and optical-MIR correlations. In the right panel, the orange lines represent the CCCDs, and CCF results appear in black.\label{ccf}}
\end{figure*}

\begin{figure*}[!htb]
\centering
\includegraphics[width=0.75\linewidth]{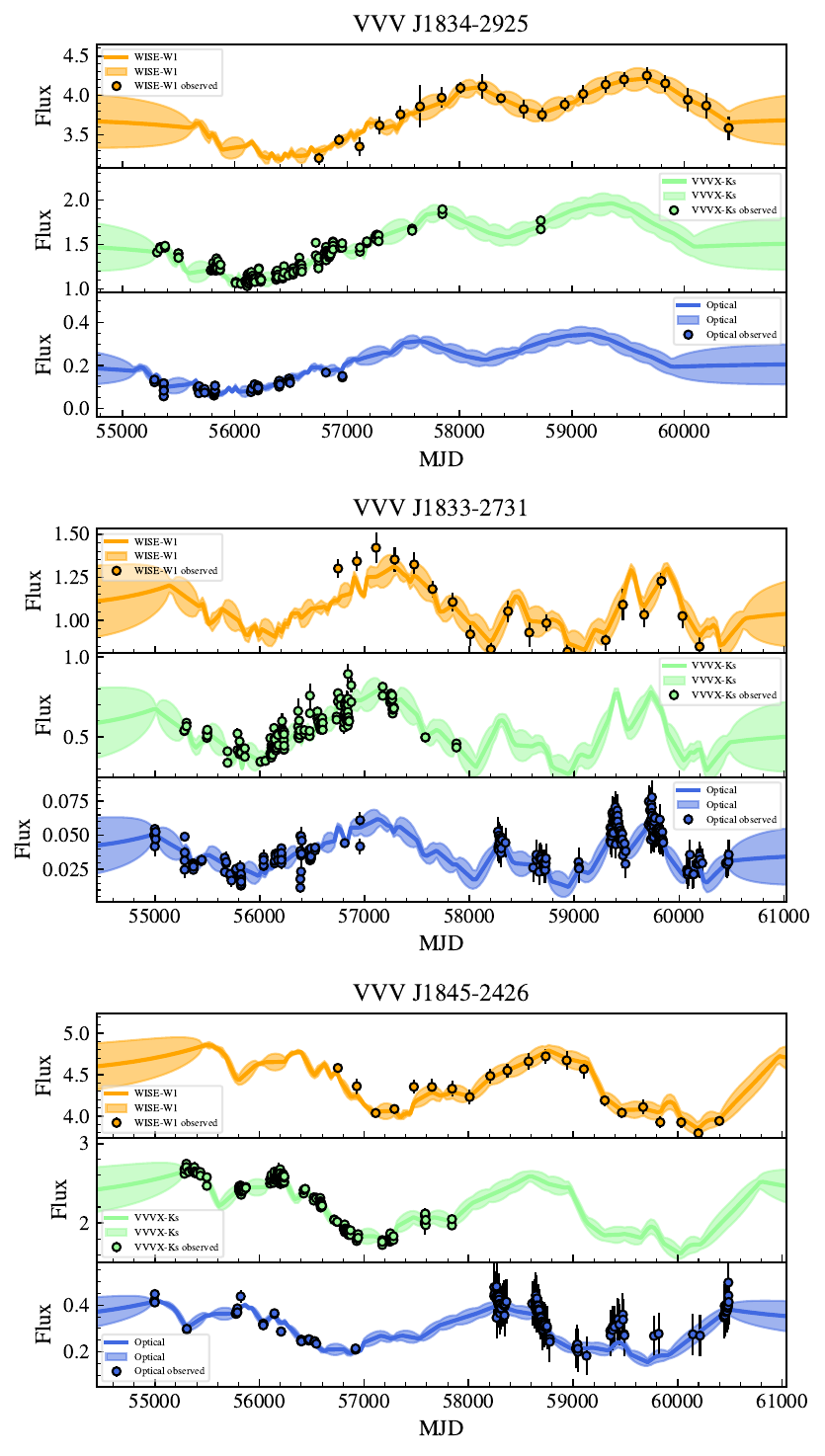}
\caption{\emph{Javelin} fits to the Optical, NIR and MIR light curves of VVV J1834-2925, VVV J1833-2731, and VVV J1845-2426.\label{drw}}
\end{figure*}

\section{Reverberation Analysis}\label{sec:analysis}
\subsection{Time Lag Analysis}\label{sec:time lags}
In Figure \ref{lc}, the quasars VVV J1833-2731 and VVV J1845-2426 exhibit excellent consistency between optical and infrared bands, with optical observations consistently preceding IR measurements. We employed the Python tool \emph{PyCCF} \citep{1998PASP..110..660P, 2018ascl.soft05032S} to estimate the time lags between light curves at different wavelengths, probing the physical relationship between optical and infrared emission regions. It determines the time lag by linearly interpolating the two light curves and computing their cross-correlation coefficient (r). To obtain robust time lags, we set the lag search range to approximately $\sim$ 70 $\%$ of the baseline of the light curves \citep{2017ApJ...851...21G, 2019ApJ...887...38G, 2020ApJ...900...58Y, 2024ApJ...968...59M}. The measured time lags are estimated based on the centroid of the interpolated cross-correlation functions (ICCFs, \citealt{1986ApJ...305..175G, 1987ApJS...65....1G}), calculated by selecting all values above $r > 0.8 r_{max}$, where $r_{max}$ corresponds to the peak of the cross-correlation function (CCF). We use the median of the distribution to represent the time lag, with the 84th and 16th percentiles indicating the upper and lower limits of its $1 \sigma$ uncertainties, respectively. The light curves and CCF results distribution are shown in Figure \ref{ccf}. It is worth noting that due to the incompleteness of the light curve data, as well as the presence of noise and errors, the CCF distribution we calculated exhibits multiple peaks. To ensure the accuracy of the results, we adopt the value corresponding to the highest peak of the cross-correlation centroid distribution (CCCD) as the estimated time lag. Due to the irregular and noisy Ks-band light curve of the target source VVV J1831-2714, which may lead to inaccurate lag time calculations, we exclude this source from the lag analysis. %Additionally, target source VVV J1834-2925 exhibits greater photometric uncertainty in W1-band than W2-band, so we calculate the CCF between the Ks-band and W2-band instead.
% \zzy{JN: For VVV J1831-2714, it seems that the optical light curve is even more noisy. Is it an optical faint, IR bright quasar?}

We also employ \emph{Javelin} \citep{2011ApJ...735...80Z,2013ApJ...765..106Z,2016ApJ...819..122Z} to determine the time delays between the optical and infrared light curves and compare the results with those obtained from the ICCF measurements. \emph{Javelin} models the light curve $X(t)$ as a DRW model and its autocovariance matrix is given by 
\begin{center}
\begin{equation}
<X(t_i)X(t_j))> = \sigma^2 exp(-|t_i-t_j|/\tau_d)
\end{equation}
\end{center}
where $t_i$ and $t_j$ are the times of two observations, and $\sigma$ and $\tau_d$ represent the amplitude scale and the exponential damping timescale, respectively. The light curve amplitude scales as $\hat{\sigma} (\tau_d/2)^{1/2}$ on long timescales, and it is $\hat{\sigma}\sqrt{t}$ on short timescales. \emph{Javelin} implements a two-phase analysis procedure. In the first phase, \emph{Javelin} characterizes the driving light curve by adopting logarithmic priors for $\tau_d$ and  $\hat{\sigma}$. Subsequently, it performs Bayesian inference through Monte Carlo Markov Chain (MCMC, \citealt{1953JChPh..21.1087M, 1970Bimka..57...97H}) sampling to jointly model N echo light curves, where Gaussian priors derived from the first-phase analysis are imposed on $\tau_d$ and $\hat{\sigma}$.

We run "Rmap" mode of \emph{Javelin} on these light curves and adopt the same search range as ICCF (-3640 to 3640 days) to identify the time lag. Due to the large lag search range applied here, multiple peaks appear in the posterior distribution of the time lag. Then we constrain the time lag range to the position of the most significant peak and re-run \emph{Javelin}. Figure \ref{drw} shows the light curves fitting obtained by \emph{Javelin}. The final time lag and its upper/lower error limits were determined from the median, 84th/16th percentiles of the probability distribution of the most prominent peak. 

The \emph{PyCCF} and \emph{Javelin} lag results are shown in Table \ref{table:lag results}. For the source VVV J1833-2731, we measure its reverberation time lags using both the \emph{PyCCF} and \emph{Javelin} methods. The results show: 1) The \emph{PyCCF} method detects aliasing effects in both Ks- and W1-band (with respect to optical band) lag measurements; 2) The \emph{Javelin} method yields an exceptionally short lag in the Ks-band (with respect to optical band). It results from the source exhibiting relatively low optical variability amplitude during modified Julian date (MJD) 55000-56000, making it difficult to reliably identify inflection points in the light curve, which consequently reduced the measurement accuracy of the Ks band (with respect to optical band) time lag.

\subsection{Redshift Effect and Accretion Disk Contamination (ADC)}\label{sec:adc}

In the preceding section, we determine the time delays of four quasars with \emph{PyCCF} and \emph{Javelin}. For all targets except VVV J1833-2731, the two methods yield consistent results within measurement uncertainties. These four sources cover a broad redshift range (0 $< z_{ph} < $1). The rest-frame time lag of the dust reverberation exhibits a positive correlation with rest-frame wavelength, meaning longer wavelengths correspond to longer delays. When observations are made at fixed bands, higher-z AGNs probe shorter rest-frame wavelengths than their lower-z counterparts. As a result, the measured rest-frame lag is inherently shorter for high-z AGNs observed in the same band. Furthermore, this rest-frame lag is cosmologically dilated by a factor of (1 + z). Therefore, to enable fair comparisons of dust reverberation sizes across AGNs spanning a broad redshift range, this wavelength–redshift coupling effect must be carefully corrected.

Followed \cite{2014ApJ...784L..11Y}, \citet{2019ApJ...886..150M} introduced a correction factor of the form $(1+z)^{\gamma}$. Using time lags between the K, H bands relative to the V band for six nearby Seyfert galaxies from the Multicolor Active Galactic Nuclei Monitoring (MAGNUM) project, they obtained $\gamma=1.18$. By combining this factor with cosmological time dilation, they derived the total correction factors as $(1+z)^{\gamma-1}=(1+z)^{0.18}$. \citet{2024ApJ...968...59M} derived the redshift correction factor $(1+z)^{\gamma-1}=(1+z)^{-0.38}$ for W1 (W2) band using a sample of 446 (416) AGNs with reliable dust reverberation measurements. We apply the total correction factors of K-band and W1 band to the sample data separately to correct for redshift effects, with the corrected time lags presented in Table \ref{table:lag results}.

The accretion disk (AD) also emits radiation in the IR band \citep{2006ApJ...652L..13T, 2008Natur.454..492K, 2011MNRAS.415.1290L}, which can cause the measured optical-IR time lag to be shorter than the actual lag. We therefore apply AD IR flux corrections to recover the true reverberation lag times. First, assuming the quasar sample adheres to the standard thin AD model, their flux density follows a power-law spectral distribution of the form: $F_{\nu} \propto \nu^{1/3}$. The IR contamination from the AD \citep{2014ApJ...788..159K} can be quantified as:
\begin{center}
\begin{equation}
F(t)_{IR, AD}=F(t)_{OP} \times (\frac{\nu_{IR}}{\nu_{OP}})^{1/3}.
\end{equation}
\end{center}
 where $F(t)_{IR, AD}$ and $F(t)_{OP}$ denote the AD's infrared and optical fluxes at time $t$, while $\nu_{IR}$ and $\nu_{OP}$ represent the effective frequencies of the infrared and optical (g) bands, respectively. The true IR emission of the dust torus is given by:
\begin{center}
\begin{equation}
F(t)_{IR, Torus} = F(t)_{IR, Observed} - F(t)_{IR, AD}.
\end{equation}
\end{center}

\begin{figure}[!ht]
\centering
\includegraphics[width=1.0\linewidth]{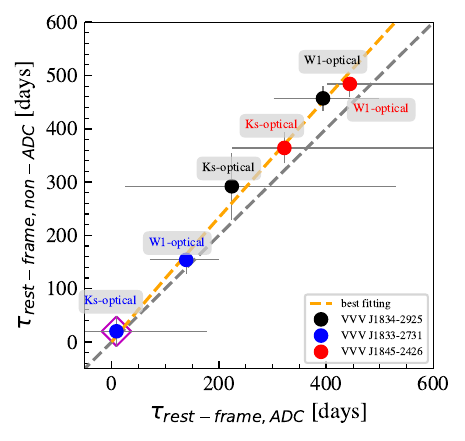}
\caption{Comparison of rest-frame time lags before (x-axis) and after (y-axis) ADC subtraction. Black, blue, and red dots correspond to sources VVV J1834-2925, VVV J1833-2731, and VVV J1845-2426, respectively. The diagonal gray dashed line indicates unity correlation. The orange line represents the best fit for all the data points except the Ks-optical lag for VVV J1833-2731 (magenta diamond dot).\label{adc}}
\end{figure}

Given the sparsely sampled light curves of our quasar sample and non-simultaneous multi-band observations, we utilize \emph{Javelin}-simulated light curves to derive the optical and infrared fluxes. We subsequently re-measure the time lags between the optical light curves and the ADC-corrected Ks and W1 light curves using \emph{PyCCF}. In Figure \ref{adc} and Table \ref{table:lag results}, we compare the redshift-corrected rest-frame with and without ADC correction time lags. The orange dashed line describes the relation between the two:
\begin{center}
\begin{equation}\label{eq:adc}
\rm log_{10}(\tau_{rest-frame, non-ADC}) \propto 0.97 log_{10}(\tau_{rest-frame, ADC}).
\end{equation}
\end{center}
The Ks-optical lag data point for J1833-2731 was discarded from the fitting owing to poor measurement reliability. The above analysis suggests that correction of ADC leads to more larger dust torus sizes.

\begin{table*}
\begin{center}
\caption{Lags estimates.
\label{table:lag results}
}
\renewcommand{\arraystretch}{1.5}
\begin{tabular}{ccccccc} 
 \hline
 \hline
 Object & Band & \multicolumn{5}{c}{Lag Time \footnote{Units: light days}} \\
 \cline{3-7}
 && \emph{PyCCF} & \emph{Javelin} & rest-frame, ADC & observed-frame, non-ADC & rest-frame, non-ADC \\
 (1) & (2) & (3) & (4) & (5) & (6) & (7)\\
 \hline
 VVV J1834-2925 & optical, Ks & $350^{+90}_{-60}$ & $200^{+270}_{-180}$ & $220^{+310}_{-200}$ & $260^{+60}_{-60}$ & $290^{+60}_{-60}$ \\
 & optical, W1 & $510^{+200}_{-320}$ & $510^{+140}_{-120}$ & $390^{+110}_{-90}$ & $590^{+30}_{-30}$ & $460^{+20}_{-20}$\\
 VVV J1833-2731 & optical, Ks \footnote{The time lags for VVV J1833-2731 between optical and Ks band are unreliable results.} & $140^{+30}_{-30}$ & $10^{+160}_{-180}$ & $10^{+170}_{-190}$ & $20^{+30}_{-30}$ & $20^{+30}_{-30}$\\
 & optical, W1 & $140^{+60}_{-60}$ & $150^{+70}_{-80}$ & $140^{+60}_{-70}$ & $170^{+10}_{-30}$ & $150^{+10}_{-30}$\\
 VVV J1845-2426 & optical, Ks & $380^{+60}_{-30}$ & $310^{+330}_{-90}$ & $320^{+340}_{-100}$ & $350^{+30}_{-30}$ & $360^{+30}_{-30}$\\
 & optical, W1 & $530^{+180}_{-150}$ & $490^{+290}_{-50}$ & $440^{+270}_{-40}$ & $530^{+10}_{-30}$ & $480^{+10}_{-30}$\\
 \hline
\end{tabular}
\end{center}
\raggedright Notes: Column (1): Name of the quasar from the VVV survey. Column (2): The two bands used for measuring dust reverberation time lags. Column (3): The observed-frame time lags derived from \emph{PyCCF}. Column (4): The observed-frame time lags derived from \emph{Javelin}. Column (5): The redshift corrected rest-frame time lags from \emph{Javelin}. Column (6): The observed-frame ADC corrected time lags using \emph{Javelin}-simulated light curves. Column (7): The rest-frame ADC corrected time lags using \emph{Javelin}-simulated light curves.
\end{table*}

\begin{figure*}[!htp]
\centering
\includegraphics[width=1.0\linewidth]{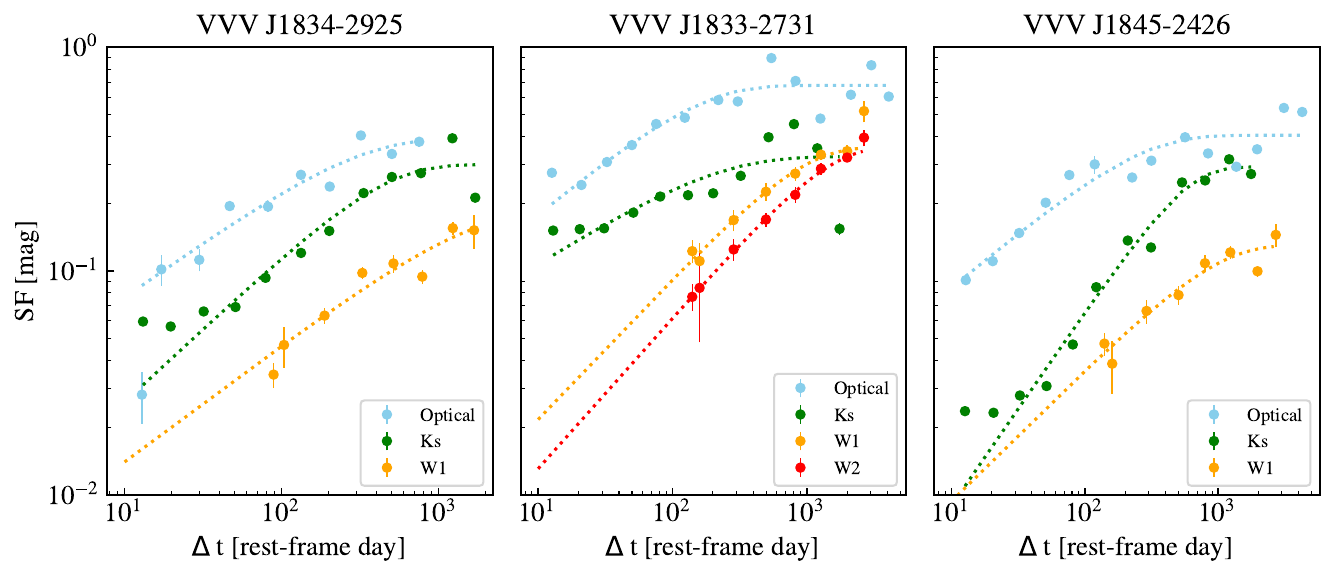}
\caption{SFs in the optical, Ks, W1 and W2 band in magnitude unit of VVV J1834-2925, VVV J1833-2731 and VVV J1845-2426. The dotted lines represents the fitting results derived from $SF(\Delta t)=SF_{\infty} \sqrt{1-e^{-(|\Delta t|/\tau)^{\beta}}}$.\label{sf}}
\end{figure*}
 
\section{Structure Function} 
\label{sec:sf}
The structure function (SF) characterizes the root mean square (rms) variability of quasars across different timescales, quantifying the magnitude or flux differences between observations separated by a rest-frame time interval $\Delta t$ ($\Delta t = \Delta t_{obs}/(1+z)$), as expressed by \citep{2023ApJ...950..122L}
\begin{center}
\begin{equation}
SF(\Delta t)=\sqrt{\frac{1}{N_{\Delta t, pairs}}\sum^{N_{\Delta t, pairs}}_{i=1}\Delta m^2}.
\end{equation}
\end{center}
The uncertainty of $SF(\Delta t)$ is determined by the standard error of the mean ($std/\sqrt{N-1}$) for all pairs of magnitude within each time interval $\Delta t$ bin. Given that the light curves from individual optical bands have limited timescales, we focus our study to the SF of the combined optical light curve. 

Figure \ref{sf} displays the SFs of three quasars in multiple bands. As shown in the Figure \ref{sf}, the optical, Ks, W1 and W2 SFs exhibit an increasing trend with $\Delta t$ at short timescales before flattening at longer timescales. Furthermore, the optical SFs exhibit larger amplitude than the W1 and W2 SFs, while the W1 and W2 SFs show steeper slopes compared to those in the optical band. This occurs because the dust torus surrounding quasars strongly absorbs and scatters optical emission, smoothing out variability signals. In contrast, 
MIR can partially penetrate the dust, more directly tracing rapid variations from the AD and dust torus—resulting in a steeper slope \citep{2023ApJ...950..122L}. The reduced MIR variability amplitudes relative to the optical arise from the extended geometry of the dusty torus, which causes the reprocessed emission to represent a time-averaged response to the central source's radiation. These results align with prior investigations \citep{2023ApJ...950..122L} of AGN SFs in both optical and MIR wavelengths. We applied $SF(\Delta t)=SF_{\infty} \sqrt{1-e^{-(|\Delta t|/\tau)^{\beta}}}$ \citep{2013ApJ...765..106Z, 2016ApJ...826..118K} to fit the optical ($\beta = 1$, DRW process), Ks, W1 and W2 SFs. The fitting results are summarized in Table \ref{table:sf}.

\begin{table*}
\begin{center}
\caption{The results of SF (in magnitude unit) fitting based on $SF(\Delta t)=SF_{\infty} \sqrt{1-e^{-(|\Delta t|/\tau)^{\beta}}}$.
\label{table:sf}
}
\renewcommand{\arraystretch}{1.5}
\begin{tabular*}{\linewidth}{@{\extracolsep{\fill}}ccccc@{}}
 \hline
 Name & Band & $SF_{\infty}$ [mag] & $\tau$ [days] & $\beta$  \\
 \cline{1-5}
 VVV J1834-2925 & Optical & $0.393^{+0.045}_{-0.045}$ & $260^{+110}_{-110}$ & $1.0$ (fixed)\\
  & Ks & $0.299^{+0.033}_{-0.033}$ & $410^{+190}_{-190}$ & $1.31^{+0.36}_{-0.36}$  \\
  & W1 & $0.183^{+0.092}_{-0.092}$ & $1300^{+2200}_{-2200}$  & $1.05^{+0.38}_{-0.38}$ \\
 \cline{1-5}
 VVV J1833-2731 & Optical & $0.677^{+0.043}_{-0.043}$ & $140^{+50}_{-50}$  & $1.0$ (fixed)\\
 & Ks & $0.325^{+0.054}_{-0.054}$ &  $160^{+190}_{-190}$ &  $0.78^{+0.48}_{-0.48}$ \\
& W1 & $0.358^{+0.016}_{-0.016}$ & $850^{+140}_{-140}$  & $1.26^{+0.11}_{-0.11}$  \\
& W2 & $0.360^{+0.021}_{-0.021}$ & $1360^{+220}_{-220}$ & $1.35^{+0.08}_{-0.08}$  \\
 \cline{1-5}
 VVV J1845-2426 & Optical & $0.419^{+0.035}_{-0.035}$ & $230^{+100}_{-100}$ & $1.0$ (fixed) \\
 & Ks & $0.299^{+0.014}_{-0.014}$ &  $570^{+90}_{-90}$ & $1.72^{+0.20}_{-0.20}$  \\
  & W1 & $0.123^{+0.011}_{-0.011}$ & $660^{+270}_{-270}$ & $1.27^{+0.45}_{-0.45}$ \\
 \hline
\end{tabular*}
\end{center}
\end{table*}

\section{Dust Torus}\label{sec:dust}
\subsection{AGN Bolometric Luminosity}\label{sec:Lbol}
\begin{figure*}[!htbp]
\centering
\includegraphics[width=0.7\linewidth]{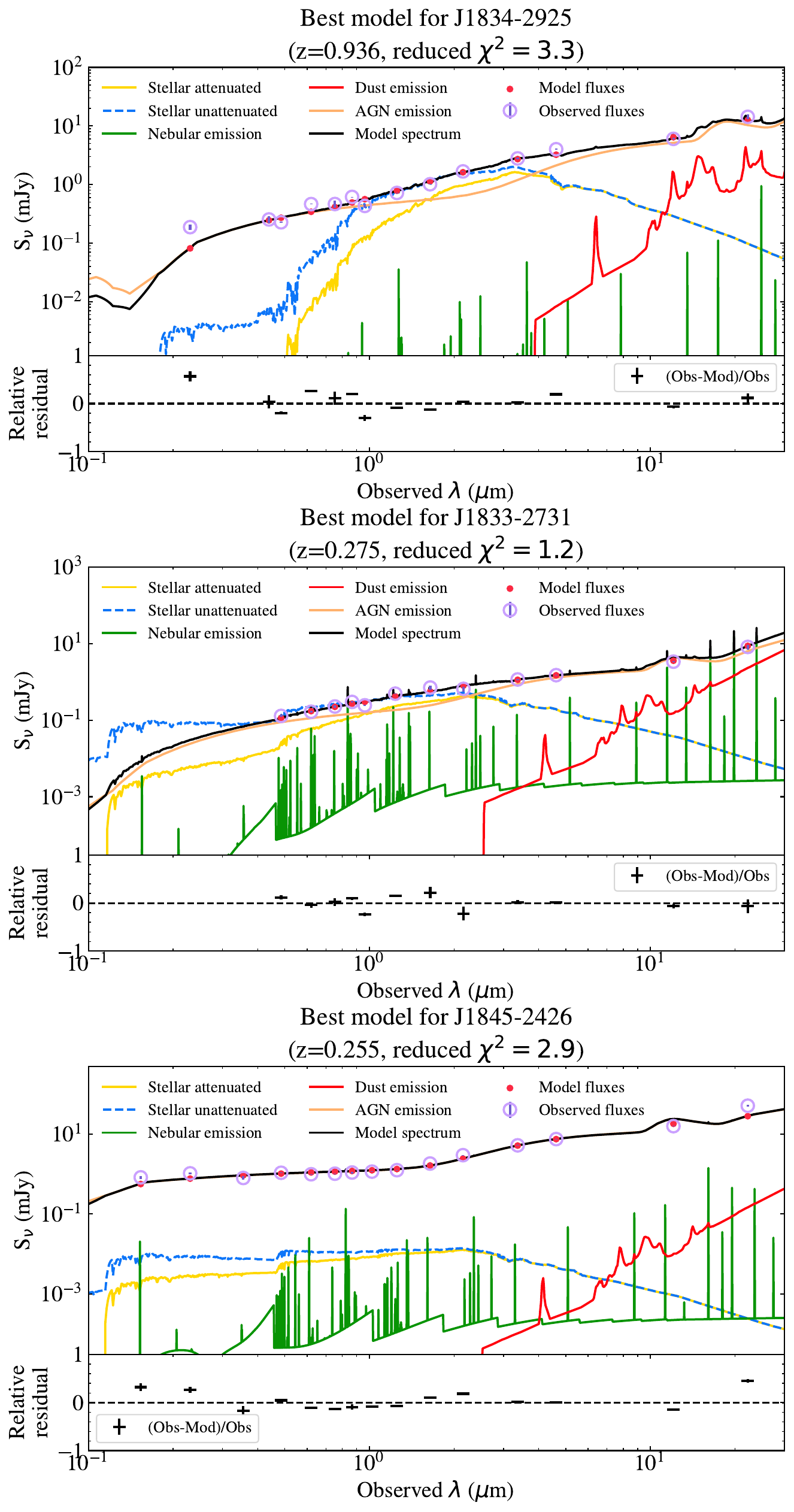}
\caption{The SED fitting of three quasars. Our CIGALE SED fitting incorporated multi-wavelength photometry spanning UV (GALEX), optical (PS1), NIR (VISTA), and MIR (WISE). The purple open and red filled circles correspond to the observed and modelled flux densities, respectively.\label{sed}}
\end{figure*}

% \textbf{Based on the photometric data from the Galaxy Evolution Explorer (GALEX), PS1, VISTA, and WISE, we perform broadband SED fitting for the three quasars in our sample using CIGALE \citep{2019A&A...622A.103B}, with the results shown in Figure \ref{sed}.} 
% %\textbf{Among the four target sources, only VVV J1845-2426 has UV photometric data. VVV J1834-2925 and VVV J1833-2731 were detected across optical to MIR bands with reliable photometric measurements, whereas VVV J1831-2714 not only lacks UV data but also has poor-quality MIR photometry.} Since this source was not included in our detailed analysis, its SED fitting results do not affect our conclusions. 
% From the SED fitting, we obtain the luminosities ($L_{AGN, SED}$) of VVV J1834-2925, VVV J1833-2731 and VVV J1845-2426 to study the AGN luminosity$-$torus size relation. \textbf{Since the primary objective of this study is to obtain the AGN luminosity through SED fitting, we utilize the CIGALE tool to perform fitting with various parameter combinations. Under the premise of ensuring reasonable physical interpretation, we ultimately select the fitting result with the smallest reduced $\chi^2$ value. Furthermore, we find that the AGN luminosities derived from different parameter combinations differed only slightly, all falling within the same order of magnitude, indicating robust and stable results. Therefore, no further physical interpretation of the SED fitting results is provided in this paper. }
Based on the photometric data from the Galaxy Evolution Explorer (GALEX), PS1, VISTA, and WISE, we perform broadband SED fitting for the three quasars in our sample using \emph{CIGALE} \citep{2019A&A...622A.103B}, with the results shown in Figure \ref{sed}. A delayed star formation history is adopted, with both the e-folding time ($\tau_{main}$) and the age ($t_{main}$) of the main stellar population allowed to vary between 1000 and 11000 Myr. The stellar component is modelled using the Bruzual-Charlot Stellar Population Synthesis model (BC03SPS) \citep{2003MNRAS.344.1000B}, combining the Chabrier initial mass function \citep{2003PASP..115..763C} and a stellar metallicity of 0.02 $Z_{\odot}$. For nebular emission lines, the color excess E(B-V) is varied between 0.01 and 0.3, and the ionization parameter is set from -4.0 to -3.0 in steps of 0.5. Given that our sample are all quasars, we add the Stalevski AGN model \citep{2016MNRAS.458.2288S}, with the AGN fraction allowed to range from 0.0 to 0.99. From the SED fitting, we obtain the luminosities ($L_{AGN, SED}$) of VVV J1834-2925, VVV J1833-2731 and VVV J1845-2426 to study the AGN luminosity$-$torus size relation.

\subsection{Torus Size and Time Lag}\label{sec:size}
The dust grains in the interstellar medium (ISM) are primarily composed of graphite and silicates, with sublimation temperature ranges of 1500-1900 K and 1000-1400 K \citep{2007A&A...476..713K}, respectively. Consequently, graphite grains can stably persist in the hottest inner regions of dust torus, while silicate grains are distributed in the cooler outer dust zones. The sublimation radii of graphite and silicate grains \citep{1987ApJ...320..537B,2015ARA&A..53..365N} can be respectively expressed as:
\setlength{\abovedisplayskip}{6pt} 
% \begin{center}
\begin{equation}\label{RC}
\begin{aligned}
\frac{R_{sub, C}}{pc} =  &1.3 (\frac{L_{UV}}{10^{46} erg s^{-1}})^{0.5} \\
& \times (\frac{T_{sub, C}}{1500 K})^{-2.8} (\frac{a_{C}}{0.05 \mu m})^{-0.5}f(\theta),
\end{aligned}
\end{equation}
% \end{center}

% \begin{center}
\begin{equation}\label{RS}
\begin{aligned}
\frac{R_{sub, S}}{pc} =  &2.7 (\frac{L_{UV}}{10^{46} erg s^{-1}})^{0.5} \\
& \times (\frac{T_{sub, S}}{1000 K})^{-2.8} (\frac{a_{S}}{0.05 \mu m})^{-0.5}f(\theta).
\end{aligned}
\end{equation}
% \end{center}
where $f(\theta)$ is the half-opening-angle-dependent term ($f(\theta)=[cos\theta(1+2cos\theta)/3]^{1/2}$) \citep{1987MNRAS.225...55N, 2013peag.book.....N} that accounts for anisotropic radiation from the AD,  $a_C$ and $a_S$ are the grain sizes of graphite and silicate \citep{2007A&A...476..713K}, respectively, and $L_{UV}$ is the UV luminosity which equals $0.165L_{AGN, bol}$ \citep{2004ASSL..308..187R}.

\citet{2021ApJ...912..126L} showed that the torus emission in 1-4$\mu$m comes from two distinct dust populations, matching the properties of sublimating graphite and silicate dust grains. The two sublimating radii correspond to the time lags of optical-Ks and optical-W1/W2, respectively. Adopting the typical sublimating temperature $T_{sub, C}$=1500K and $T_{sub, S}$=1000K for graphite and silicate, we derive the ratio of grain size $\frac{a_C}{a_S} \sim$ 0.4 for J1845-2426 from Equations (\ref{RC}) and (\ref{RS}). This result demonstrates that small graphite grains dominate the NIR emission, while large silicate grains primarily contribute to MIR radiation. This agrees well with the dust model presented by \citet{2007ApJ...657..810D}, where NIR (short wavelength) emission is strongly produced by the smallest dust particles and it can be heated to T $>$ 1000K through absorption of a single UV photon. NIR emission efficiency decreases significantly with increasing grain size
and dust-absorbed energy is progressively re-radiated in the MIR.

\subsection{The Relation Between Torus Size and Luminosity}\label{sec:sizevsLum}
Below we use the time lag between the optical and Ks/W1 bands as a proxy for the torus size. We explore the connection between torus size and AGN luminosity using a linear relationship:
 \begin{center}
\begin{equation}\label{eq:RL}
\rm log(\frac{R_{torus}}{light \ day}) = \alpha + \beta log(\frac{L_{AGN}}{L_0}).
\end{equation}
\end{center}

Note that we adopt the time lags with ADC correction and redshift correction (Sec.\ref{sec:adc}) in this section. Due to the limited number of quasars in our sample, we supplement it with existing NIR \citep{2014ApJ...788..159K, 2019ApJ...886..150M} and MIR \citep{2019ApJ...886...33L, 2024ApJ...968...59M} RM AGNs from the literature. We then apply the updated quasar luminosity bolometric correction relation from \citet{2012MNRAS.426.2677R} to derive $L_{AGN, bol}$ from $\lambda L_{\lambda}(0.51 \mu m)$ for NIR RM sample \citep{2000ApJ...533..631K, 2014ApJ...788..159K, 2019ApJ...886..150M}
\begin{center}
\begin{equation}
\rm log(\frac{L_{AGN, bol}}{erg \ s^{-1}})=4.89+0.91log(\frac{\lambda L_{\lambda}(0.51 \mu m)}{erg \ s^{-1}}).
\end{equation}
\end{center}
For the MIR RM sample, we exclude AGNs with peak CCF $<$ 0.8 from \citet{2024ApJ...968...59M}’s sample, as well as those from \citet{2019ApJ...886...33L}’s sample exhibiting large uncertainties, weak variability features, or peak CCF $<$ 0.7.

\begin{figure*}[!htbp]
\centering
\includegraphics[width=1.0\linewidth]{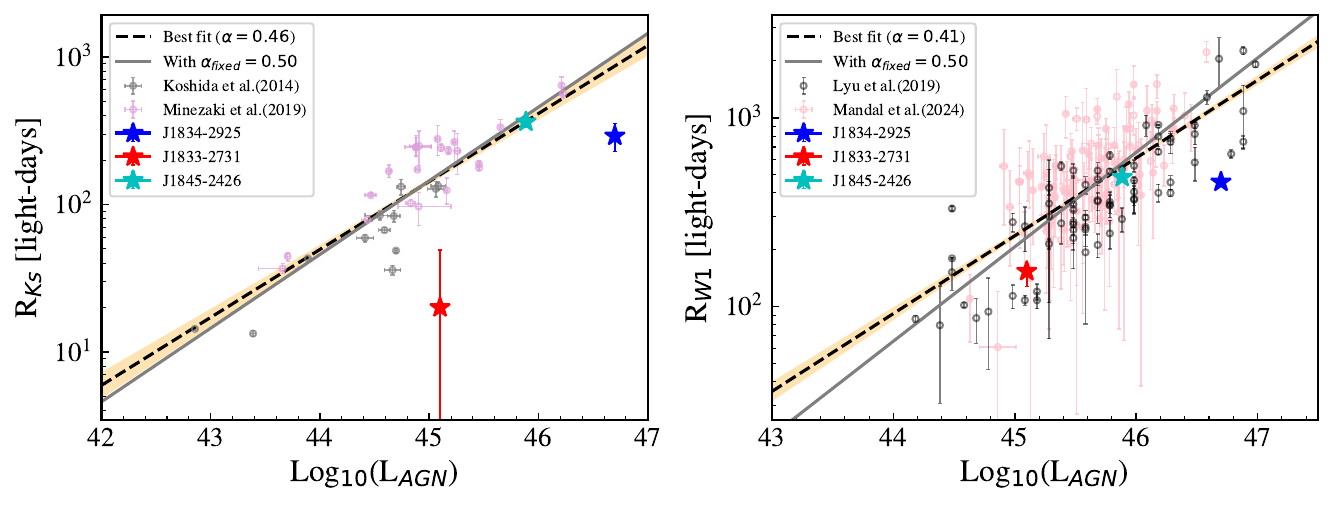}
\caption{The dust torus size-AGN luminosity relation in Ks (left) and W1 (right) bands after ADC and redshift corrections. The three quasar samples of this work are represented by solid pentagrams in blue, red, and cyan. Left column: the gray and purple open circles represent AGN samples from \citet{2014ApJ...788..159K} and \citet{2019ApJ...886..150M}. Right column: the black and pink open circles denote AGN samples from \citet{2019ApJ...886...33L} and \citet{2024ApJ...968...59M}. The black dashed line indicates the best fit to the data points (excluding J1833-2731 in the left panel), with the orange shaded region showing the 3$\sigma$ uncertainty range. For comparison, the gray solid line displays the best fit result with a fixed slope of 0.5.\label{L-R}}
\end{figure*}

Figure \ref{L-R} shows the distribution of all samples used in this study on the $L_{AGN}$-$R_{Ks/W1}$ plane. We fit Eq.(\ref{eq:RL}) using the Bivariate Correlated Errors and intrinsic Scatter (BCE, \citealt{1996ApJ...470..706A}) method, which incorporates measurement errors and correlations for both variables. The best fits give $\alpha=0.77_{-0.14}^{+0.14}$, $\beta=0.46_{-0.05}^{+0.05}$ for Ks band and $\alpha=1.96_{-0.06}^{+0.06}$, $\beta=0.41_{-0.03}^{+0.03}$ for W1 band. Since the time lag of J1833-2731 has large uncertainty, we exclude it from the $L_{AGN}$-$R_{Ks}$ fit but displayed it in Figure \ref{L-R} (left panel) for reference. Our fitting results show that the Ks band relation closely follows the $R \propto L^{0.5}$ dependence, while shallower slope is observed in the W1 band. According to the classical model by \citet{1987ApJ...320..537B}, when dust grains in the torus reach thermal equilibrium with the 
IR re-emission heated by the central AGN's UV radiation, the dust radius should follow the relation $R_{dust, IR} \propto L_{IR}^{0.5}$. \citet{2023MNRAS.522.3439C} and \citet{2024ApJ...968...59M} suggested that the shallow slopes observed may result from the self-shadowing effects of the slim disk \citep{2011ApJ...737..105K, 2014ApJ...797...65W, 2018ApJ...856....6D}, which would lead to an apparent reduction in the observed torus radius and consequently flatten the $R_{dust}-L_{bol}$ relation. This effect was not detected in the Ks band, likely because the thirteen sources in the sample of \cite{2014ApJ...788..159K} have Eddington ratios too small to form the slim disk.

\subsection{Polar Dust}
We test the existence of polar dust in VVV J1834-2426, which is classified as a Type 1 AGN because of its broad-line spectral feature. 
 %Based on the spectral characteristics of VVV J1845-2426, this object is classified as a Type 1 AGN. 
 However, the absence of distinct absorption lines in its spectrum prevents us to confirm the existence of the polar dust. %confirmation of the presence of polar dust. 
 We further perform the SED fitting analysis with the semi-empirical SED library established by \citet{2018ApJ...866...92L}, which is constructed based on Type 1 AGNs spanning a wide range of luminosities ($L_{bol} \sim 10^8-10^{14} L_{\odot}$) and redshifts ($z \sim 0-6$), with an additional polar dust component. In this library, there are three fundamental AGN types, the normal AGNs described by the Elvis-like template, the warm-dust-deficient (WDD) AGNs, and the hot-dust-deficient (HDD) AGNs \citep{2017ApJ...841...76L}.

\begin{figure}[!htbp]
\centering
\includegraphics[width=1.0\linewidth]{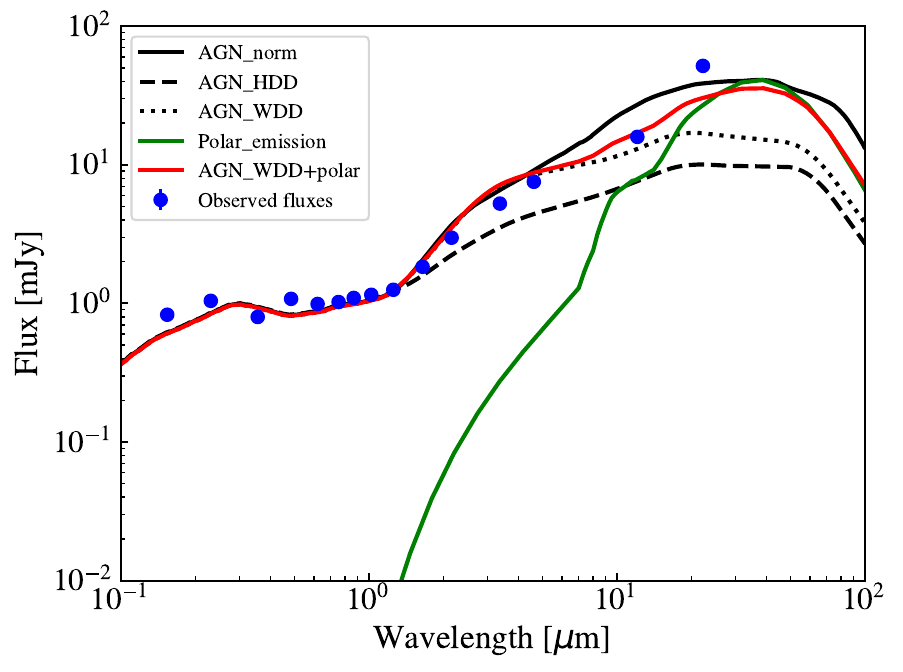}
\caption{Fitting results for VVV J1845-2426 with the AGN intrinsic templates (norm: black solid line, WDD: black dotted line, and HDD: black dashed line) and polar emission component (green line). The best fitting template (red line) is the combination of WDD AGNs and polar dust emission.\label{polar dust}}
\end{figure}

We fit the photometric data of VVV J1845-2426 using the intrinsic AGN templates (normal, WDD, and HDD) and the polar dust emission template provided by \citet{2018ApJ...866...92L}, which are shown in the Figure \ref{polar dust}. %The figure shows that all three basic templates, especially the HDD AGNs template, provide a poor fit to the photometric data of VVV J1845-2426 without polar dust. 
A simple combination of the WDD AGNs template and polar dust emission led to a significant improvement in the fitting.
%Despite the limitation imposed by the lack of MIR and FIR data, which precludes a more precise SED analysis, 
The improved fit suggests the presence of polar dust in this source, although there is a lack of the FIR data.

\begin{figure}[!htbp]
\centering
\includegraphics[width=1.0\linewidth]{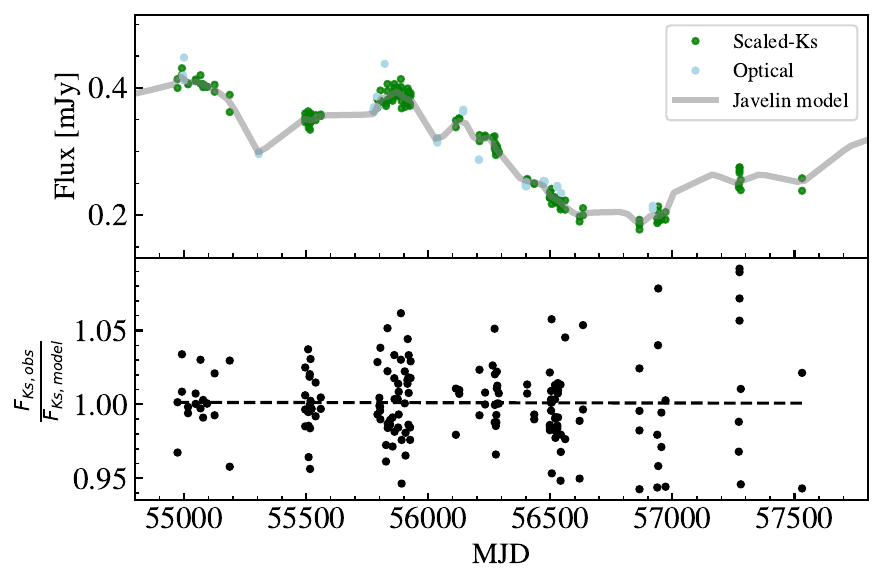}
\caption{Top panel: the comparison of optical band (light blue dots) and lag-corrected, scaled Ks-band (green dots) light curves of VVV J1845-2426. The gray solid line represents the best-fitting \emph{Javelin} model for Ks-band light curve. Bottom panel: the ratio of Ks-band observed flux to the predicted Ks-band flux by the best-fitting model as a function of time (MJD).  \label{Ks-optical}}
\end{figure}

Polar dust originates from dusty outflows at the inner edge of the circumnuclear dust torus. Driven by radiation pressure, thermal mechanisms, or other processes, gas and dust are lifted vertically, forming a dusty wind \citep{2012ApJ...755..149H, 2013ApJ...771...87H}. If such a wind persists, the torus would expose more dust, leading to a gradual increase in Ks-band flux over time \citep{2021ApJ...912..126L}. However, we don't find such trend in the Figure \ref{Ks-optical}. 
%The top panel of Figure \ref{Ks-optical} compares the optical light curve with the lag-corrected and scaled Ks-band light curve, where the solid gray line represents the best-fit model for the Ks band. The bottom panel of Figure \ref{Ks-optical} shows the ratio between the observed Ks-band data and the model predictions. 
Between MJD 55000 and 57000, the optical and Ks-band light curves are highly consistent, indicating that the variations in the Ks band are entirely due to the reprocessing of optical radiation by the inner dust torus. To extract the systematic trend of Ks-band, we perform a \emph{Javelin} model to fit its light curve, as shown in the top panel of Figure \ref{Ks-optical}. The best-fit model accurately predicts the Ks-band variability, with no discernible trend of the flux deviating from the model and increasing over time.

SED fitting (Figure \ref{polar dust}) confirms the presence of polar dust, while the variablity of Ks-band may indicate no dusty outflow in VVV J1845-2426. A possible explanation for this apparent contradiction is that the dusty outflow ceases, and the inner torus is in a relatively stable state now. The polar dust may have formed from dusty outflows that occurred in the distant past.

\section{Conclusion} \label{sec:highlight}
We present a variability study of four quasars using photometric data from PS1, ZTF, VVV/VVVX, and WISE. By applying RM techniques, we measure inter-band time lags and connect them to dust torus scales. By combining our measurements with archival AGN samples, we further explore the relationship between dust torus sizes and AGN luminosity. Our key findings are as follows:
\begin{itemize}[label=\linenumbers] % 实心圆点(默认)
    \item [1.] By cross-matching the CatNorth and VVV/VVVX surveys, we select four quasars with Ks-band variability amplitudes $>$ 0.5 mag as our primary targets for this study. By fitting the spectrum of VVV J1845-2426 with \emph{QSOFITMORE}, we determine its black hole mass to be $\log(M_{\rm BH}/M_\odot) = 8.6^{+0.1}_{-0.1}$.
    \item [2.] After correcting the infrared flux contamination from accretion disk and redshift effect, we measure inter-band time lags for the three quasars VVV J1834-2925, VVV J1833-2731, and VVV J1845-2426, as listed in Table \ref{table:lag results}. The rest-frame ADC-corrected torus sizes are systematically larger than their rest-frame uncorrected counterparts.
    \item [3.] We measure the structure functions of the three quasars across optical, Ks, W1, and W2 bands. The structure functions initially rise with $\Delta t$ on short timescales before flattening at longer timescales. Notably, the MIR structure functions exhibit both smaller variability amplitudes (in magnitude unit) and steeper slopes compared to their optical counterparts.
    \item[4.] Based on the theoretical formulation of sublimation radii for graphite and silicate, we combine reverberation time lags and sublimation temperatures to derive the graphite-to-silicate grain size ratio $\frac{a_C}{a_S}\sim$ 0.4. The result indicates that small graphite grains dominate NIR emission, whereas larger silicate grains primarily contribute to MIR radiation.
    \item [5.] We investigate the dust torus size-luminosity relation for VVV J1833-2731, VVV J834-2925, and VVV J1845-2426, and find that it is consistent with the established $R_{dust}-L_{AGN}$ relations reported by \citet{2014ApJ...788..159K} (K-band), \citet{2019ApJ...886..150M} (K-band), \citet{2019ApJ...886...33L} (W1 band), and \citet{2024ApJ...968...59M} (W1 band.)
    \item [6.] We discuss the polar dust properties of VVV J1845-2426. The SED fitting results suggest the existence of polar dust in this source.
\end{itemize}

The AGN accretion disk-to-dust torus system typically spans $<$ 1 pc, a scale too small for direct imaging with current telescopes. RM circumvents this limitation by translating spatial structure into wavelength-dependent time lags. However, while time-domain surveys have advanced rapidly, comprehensive multi-wavelength RM studies $-$ particularly those simultaneously covering optical, NIR, and MIR bands $-$ remain rare. While the studies by \citet{2018rnls.confE..57S} and \citet{2021ApJ...912..126L} similarly focused on a detailed multi-wavelength analysis of specific variable AGN sources, our study differs in its initial selection method. Rather than studying pre-known variable AGNs, we systematically select our targets from the large-scale NIR time-domain VVV survey by cross-matching with the synthetic quasar catalog CatNorth and requiring high variability amplitude. This selection strategy represents a step towards systematically identifying such variable targets for future in-depth studies with upcoming surveys.
%In contrast, our study utilizes the large-scale NIR time-domain VVV survey and synthetic quasar candidates CatNorth compiled from multi-survey data to select AGN based on variability amplitude, enabling a systematic analysis of variability propertites and multi-wavelength RM, rather than being limited to preselected targets.

\begin{acknowledgments}
We would like to thank the anonymous referee for valuable comments. This work is supported by National Key R\&D Program of China No.2022YFF0503402. We also acknowledge the science research grants from the China Manned Space Project, especially, NO.  CMS-CSST-2025-A18. ZYZ acknowledges the support by the China-Chile Joint Research Fund (CCJRF No. 1906).

We acknowledge the support of the staff of the Lijiang 2.4 m telescope. Funding for the telescope has been provided by Chinese Academy of Sciences and the People's Government of Yunnan Province.

We gratefully acknowledge the use of data from the ESO Public Survey program IDs 179.B-2002 and 198.B2004 taken with the VISTA telescope and data products from the Cambridge Astronomical Survey Unit (CASU) and the VISTA Science Archive (VSA) and the ESO Science Archive.

This work uses data from the Pan-STARRS1 Surveys. The PS1 and the PS1 public science archive have been made possible through contributions by the Institute for Astronomy, the University of Hawaii, the Pan-STARRS Project Office, the Max-Planck Society and its participating institutes, the Max Planck Institute for Astronomy, Heidelberg and the Max Planck Institute for Extraterrestrial Physics, Garching, The Johns Hopkins University, Durham University, the University of Edinburgh, the Queen's University Belfast, the Harvard-Smithsonian Center for Astrophysics, the Las Cumbres Observatory Global Telescope Network Incorporated, the National Central University of Taiwan, the Space Telescope Science Institute, the National Aeronautics and Space Administration under grant No. NNX08AR22G issued through the Planetary Science Division of the NASA Science Mission Directorate, the National Science Foundation grant No. AST-1238877, the University of Maryland, Eotvos Lorand University (ELTE), the Los Alamos National Laboratory, and the Gordon and Betty Moore Foundation.

This work is based on observations obtained with the Samuel Oschin Telescope 48-inch and the 60-inch Telescope at the Palomar Observatory as part of the Zwicky Transient Facility project. ZTF is supported by the National Science Foundation under Grants No. AST1440341 and AST-2034437 and a collaboration including current partners Caltech, IPAC, the Weizmann Institute for Science, the Oskar Klein Center at Stockholm University, the University of Maryland, Deutsches Elektronen-Synchrotron and Humboldt University, the TANGO Consortium of Taiwan, the University of Wisconsin at Milwaukee, Trinity College Dublin, Lawrence Livermore National Laboratories, IN2P3, University of Warwick, Ruhr University Bochum, Northwestern University and former partners the University of Washington, Los Alamos National Laboratories, and Lawrence Berkeley National Laboratories.

This work makes use of data products from the Wide-field Infrared Survey Explorer, which is a joint project of the University of California, Los Angeles, and the Jet Propulsion Laboratory/California Institute of Technology, funded by the National Aeronautics and Space Administration. This work also makes use of data products from NEOWISE-R, which is a project of the Jet Propulsion Laboratory/California Institute of Technology, funded by the Planetary Science Division of the National Aeronautics and Space Administration.

\end{acknowledgments}

% \begin{contribution}
%%This section gives authors the space to recognize author contributions. The text inside this environment is NOT counted towards the total word quanta. At a minimum, manuscripts are expected to include this text:

% All authors contributed equally to the Terra Mater collaboration.

%% But authors are expected to provide more specific details, e.g. 
%%
%%SC was responsible for writing and submitting the manuscript.
%%WWM came up with the initial research concept and edited the manuscript.
%%OTS obtained the funding and edited the manuscript.
%%EBF provided the formal analysis and validation. He also edited the manuscript.
%%GEH Supervised the undergraduates, wrote the software and administers the project github and Zenodo repositories.
%%
%% Authors can use the Contributor Role Taxonomy (CRediT) at
%% for ideas on how write a good statement tailored to their needs.

% \end{contribution}

\facilities{VISTA(VVV/VVVX), PS1, ZTF, WISE(NEOWISE), LJT}

\software{astropy \citep{2013A&A...558A..33A,2018AJ....156..123A,2022ApJ...935..167A}, PyCALI \citep{2024zndo..10700132L}, PyFOSC \citep{yuming_fu_2024_10967240}, PyCCF \citep{1998PASP..110..660P}, QSOFITMORE \citep{2021zndo...5810042F}, Javelin \citep{2011ApJ...735...80Z, 2013ApJ...765..106Z, 2016ApJ...819..122Z}, CIGALE \citep{2019A&A...622A.103B}, BCE \citep{1996ApJ...470..706A}, TOPCAT \citep{2005ASPC..347...29T}}


% \appendix

% \section{Appendix information}

% \section{Author publication charges} \label{sec:pubcharge}

% \section{Rotating tables} \label{sec:rotate}

% \section{Using Chinese, Japanese, and Korean characters}


\bibliography{sample7}{}
\bibliographystyle{aasjournal}


\end{CJK*}
\end{document}